\documentclass[aps,prl,reprint,superscriptaddress,nofootinbib]{revtex4-2}

\usepackage{amsmath,amssymb,bm}
\usepackage{graphicx}
\usepackage[colorlinks=true,citecolor=blue,linkcolor=blue,urlcolor=blue]{hyperref}
 
\begin{document} 

\title{Acceleration-induced spectral blind spots in stimulated atomic transitions}

\author{Jiawei Hu}
\author{Hongwei Yu}
\email{hwyu@hunnu.edu.cn}
\affiliation{Department of Physics, Key Laboratory of Low Dimensional Quantum Structures and Quantum Control of Ministry of Education, and Hunan Research Center of the Basic Discipline for Quantum Effects and Quantum Technologies, Hunan Normal University, Changsha, Hunan 410081, China}

%\date{\today}

\begin{abstract}

Stimulated transitions are among the most fundamental processes in light-matter interaction, underlying resonant absorption and emission in atomic systems. Here we show that uniform acceleration can convert this familiar response into a frequency-selective absence of response. Specifically, when an incident photon has a nonzero momentum component transverse to the acceleration, the stimulated transition probability vanishes at a discrete set of frequencies fixed by the acceleration, the atomic transition frequency, and the photon propagation angle. At these spectral blind spots, both ordinary stimulated absorption and acceleration-induced excitation are simultaneously suppressed, rendering the atom effectively unresponsive to the incident radiation. The effect arises from the nontrivial response of accelerated atoms to quantum vacuum fluctuations  and provides a distinctive signature of the Unruh effect through the absence, rather than the enhancement, of stimulated transitions. We further provide an order-of-magnitude estimate showing that an electron-based implementation with spin splitting in combined electric and magnetic fields could access the required parameter regime. These results reveal an unexplored form of acceleration-modified light-matter interaction and identify spectral blind spots as a new manifestation of the Unruh effect.

\end{abstract}

\maketitle

\emph{Introduction}---Resonant light--matter interaction is governed by a simple principle: an atom responds when the incident field supplies the energy required for a transition. This principle underlies absorption spectroscopy, stimulated emission, cavity quantum electrodynamics, and laser-driven control of quantum emitters.  Inertial motion can shift the resonance condition through the relativistic Doppler effect, but resonant photons are generally expected to enhance, rather than suppress, atomic transitions. A natural question is therefore whether acceleration can fundamentally reshape the stimulated response of an atom.

Here we show that it can. A uniformly accelerated two-level atom can exhibit \emph{spectral blind spots} in stimulated atomic transitions: discrete frequencies at which the stimulated transition probability vanishes despite the presence of incident photons. These blind spots occur when the incident photon has a nonzero momentum component transverse to the acceleration, and their locations are determined jointly by the acceleration, the atomic transition frequency, and the photon propagation angle. As a result, the stimulated response of an accelerated atom is  punctuated by sharply defined frequencies at which the atom becomes effectively unresponsive to the incident radiation.

The physical origin of this phenomenon lies in the nontrivial response of accelerated quantum systems to vacuum fluctuations. Uniform acceleration occupies a special place  in relativistic quantum theory because of the Unruh effect, which predicts that a uniformly accelerated observer perceives the Minkowski vacuum as a thermal bath with a temperature proportional to the proper acceleration \cite{Fulling73,Davies75,unruh1976notes}.  Closely related to Hawking radiation from black holes \cite{Hawking74,Hawking75}, the Unruh effect remains extremely difficult to observe directly, since measurable temperatures require enormous accelerations. For example, an Unruh temperature of $1\,\mathrm{K}$ corresponds to an acceleration of order $10^{20}\,\mathrm{m/s^2}$.

Consequently, a wide variety of proposals have been developed to search for signatures of acceleration-modified vacuum fluctuations, such as spontaneous excitation of accelerated detectors, acceleration radiation from electrons, cavity-enhanced emission, geometric phases, and collective amplification \cite{Pisin99,Habs06,Scully03,Lochan22,Martin11,Hu12,Lochan25}. Related noninertial effects have also been studied in uniformly rotating systems \cite{Bell83,Bell87,Unruh98,Rogers88,Lochan20,Arya2022,Arya2023,Zhou25a,Zhou25b} and in analog platforms such as Bose--Einstein condensates and superconducting Josephson junctions \cite{Retzker2008,Gooding2020,Katayama2025}. From a quantum-optical perspective, these studies have largely focused on how acceleration generates, enhances or modifies otherwise weak vacuum-induced processes. By contrast, the spectral blind spots identified here arise from the suppression of stimulated transitions and therefore represent a qualitatively different manifestation of acceleration-induced quantum effects.

To demonstrate this effect, we consider a uniformly accelerated two-level atom irradiated by external photons. Incident photons can drive two stimulated  excitation channels: ordinary stimulated absorption and stimulated Unruh excitation, corresponding respectively to atomic excitation accompanied by photon absorption and by photon emission. We show that, for photons propagating neither parallel nor antiparallel to the acceleration, both transition amplitudes are governed by the same modified Bessel function. At the zeros of this function, ordinary stimulated absorption and stimulated Unruh excitation vanish simultaneously, rendering the accelerated atom effectively blind to the incident radiation. This behavior contrasts sharply with that of an inertial atom, whose response is governed  solely by usual resonance condition. Under uniform acceleration, the stimulated response instead develops discrete blind-spot frequencies determined by the acceleration, the atomic level spacing, and the transverse photon momentum. We further show that these blind spots persist under coherent driving, enabling experimental tests with laser or microwave fields without resolving or postselecting the final radiation state. Finally, we discuss an electron-based implementation in which the acceleration and spin splitting can be independently controlled by external electric and magnetic fields.

Throughout this work, we adopt natural units with $c = \hbar = k_B = 1$, where $c$ is the speed of light, $\hbar$ is the reduced Planck constant, and $k_B$ is the Boltzmann constant.

\emph{Driven two-level atom along an accelerated trajectory}---We first formulate the driven transition amplitudes for a two-level atom weakly coupled to a massless scalar field $\phi(\boldsymbol{x}, t)$. The ground and excited states of the atom are denoted by $|g\rangle$ and $|e\rangle$, respectively, with an energy level spacing $\omega_0$. In the interaction picture, the interaction Hamiltonian takes the form
\begin{equation}
 H_{I}(\tau)= \lambda \mu(\tau) \phi(\boldsymbol{x}(\tau), t(\tau)),
\end{equation}
where $\lambda$ is the coupling strength,  $\mu(\tau)=e^{i \omega_0 \tau} \sigma_{+} + e^{-i \omega_0 \tau} \sigma_{-}$ is the atomic monopole moment operator, and 
\begin{eqnarray}
\phi(\boldsymbol{x}(\tau), t(\tau))&=&\int \frac{d^{3} k}{(2 \pi)^{3}} \frac{1}{\sqrt{2 \omega_{\boldsymbol{k}}}}(e^{-i \omega_{\boldsymbol{k}} t(\tau)+i \boldsymbol{k} \cdot \boldsymbol{x}(\tau)} a_{\boldsymbol{k}}\nonumber\\
&&+e^{i \omega_{\boldsymbol{k}} t(\tau)-i \boldsymbol{k} \cdot \boldsymbol{x}(\tau)} a_{\boldsymbol{k}}^{\dagger})
\end{eqnarray}
is the quantum field evaluated along the atomic trajectory $x^\mu(\tau)=(\boldsymbol{x}(\tau), t(\tau))$, with $\tau$ being the  atom's proper time. In the equations above, $\omega_{\boldsymbol{k}}$ is the frequency of the field mode, and $a_{\boldsymbol{k}}$ ($a_{\boldsymbol{k}}^{\dagger}$) are the annihilation (creation) operators satisfying the standard commutation relations.
The atomic raising and lowering operators are $\sigma_+ = |e\rangle\langle g|$ and $\sigma_- = |g\rangle\langle e|$.

The transition amplitude of the combined atom--field system  from an initial state $|\psi_{\alpha}\rangle$ to a final state $|\psi_{\beta}\rangle$ is
\begin{equation}\label{3}
    \mathcal{A}_{\alpha \rightarrow \beta}
    =\int\left\langle\psi_{\beta}\left|H_{I}(\tau)\right| \psi_{\alpha}\right\rangle d \tau=\int \frac{d^{3} k}{(2 \pi)^{3}} \frac{\mathcal{A}_{\alpha \rightarrow \beta}(\boldsymbol{k})}{\sqrt{2 \omega_{\boldsymbol{k}}}} ,
\end{equation}
where the contribution from the field mode with momentum $\boldsymbol{k}$ is
\begin{eqnarray}
    \mathcal{A}_{\alpha \rightarrow \beta}(\boldsymbol{k})&=&
    \lambda [I_{-}(\boldsymbol{k})\langle\psi_{\beta}|\sigma_{+} a_{\boldsymbol{k}}| \psi_{\alpha}\rangle\nonumber\\
    &&+I_{-}^*(\boldsymbol{k})\langle\psi_{\beta}|\sigma_{-} a_{\boldsymbol{k}}^{\dagger}| \psi_{\alpha}\rangle\nonumber\\
    &&+I_{+}(\boldsymbol{k})\langle\psi_{\beta}|\sigma_{+} a_{\boldsymbol{k}}^{\dagger}| \psi_{\alpha}\rangle\nonumber\\
    &&+I_{+}^*(\boldsymbol{k})\langle\psi_{\beta}|\sigma_{-} a_{\boldsymbol{k}}| \psi_{\alpha}\rangle]
\end{eqnarray}
 with
\begin{equation}\label{I}
    I_{\pm}(\boldsymbol{k})=\int d \tau\, e^{i \omega_0 \tau \pm i k^{\mu} x_{\mu}(\tau)}.
\end{equation}
The quantities $I_-(\mathbf{k})$ and $I_+(\mathbf{k})$ characterize two distinct excitation channels. The first corresponds to ordinary stimulated absorption, in which the atom is excited while absorbing a photon. The second corresponds to stimulated Unruh excitation, in which the atom is excited while emitting an additional photon. The corresponding probabilities are
\begin{eqnarray}\label{resonant-ab}
    &&\left|\mathcal{A}_{|g, n\rangle \rightarrow |e, n-1\rangle}(\boldsymbol{k})\right|^2=n\lambda^2 \left|I_{-}(\boldsymbol{k})\right|^2,\\
    &&\left|\mathcal{A}_{|g, n\rangle \rightarrow |e, n+1\rangle}(\boldsymbol{k})\right|^2=(n+1)\lambda^2 \left|I_{+}(\boldsymbol{k})\right|^2,\label{Unruh-ex}
\end{eqnarray}
where $\left|n\right\rangle=\frac{1}{\sqrt{n !}} a_{\boldsymbol{k}}^{\dagger}|0\rangle$ is the Fock state containing $n$ photons in mode $\boldsymbol{k}$.  
In what follows, we investigate how these driven transition processes behave when the atom undergoes uniform acceleration.

\emph{Acceleration-induced spectral blind spots}---We now evaluate the stimulated transition amplitudes for a uniformly accelerated two-level atom following the trajectory
\begin{equation}\label{tra}
t(\tau)=\frac{1}{a}\sinh{a\tau},~x(\tau)=\frac{1}{a}\cosh{a\tau},~y(\tau)=z(\tau)=0,
\end{equation}
where $a$ is the proper acceleration. We consider the contribution from a field mode with four-momentum  $k_\mu=(k_{0},k_\parallel,k_\perp,0)$, where $k_0$ is the photon frequency, and $k_\parallel$ and $k_\perp$ are the momentum components parallel and transverse to the acceleration, respectively. 
Substituting the trajectory Eq. \eqref{tra} into Eq. \eqref{I} gives 
\begin{eqnarray}\label{i1}
    I_{\pm}(\boldsymbol{k})=
    \int_{-\infty}^{\infty}d \tau\, e^{i\omega_0 \tau \pm \frac{i}{a}(k_{0}\sinh{a\tau}-k_{\parallel}\cosh{a\tau})}.
\end{eqnarray}

%%%%%%%%%%%%%%%%%%%%%%%%%%%%%%%%%%%%%%%%%%%%%%%%%%

\emph{A. Oblique driving fields}---We first consider photons with a nonzero transverse momentum component, $k_\perp\neq 0$, such that the incident photon propagates neither parallel nor antiparallel to the acceleration.

When $k_\parallel>0$, introducing the parameter $\zeta$ via $\cosh\zeta = k_0/k_\parallel$ allows Eq.~\eqref{i1} to be rewritten as 
\begin{eqnarray}\label{i3}
    I_{\pm}(\boldsymbol{k})=
    \int_{-\infty}^{\infty}d \tau\, e^{i\omega_0 \tau \pm i\frac{k_{\perp}}{a} \sinh{(a\tau-\zeta)}}.
\end{eqnarray}
After shifting the integration variable to $\tau'=\tau-\zeta/a$ and using Eq.~(3.996) of Ref.~\cite{gradshteyn2014table}, we obtain
\begin{eqnarray}\label{i3-1}
       I_{\pm}(\boldsymbol{k})=
    \frac{2}{a}\, e^{i\frac{\omega_0\zeta}{a}} e^{\mp\frac{\pi\omega_0}{2a}} K_{{i\omega_0}/{a}}\left(\frac{k_\perp}{a}\right),
\end{eqnarray}
where $K_{i\nu}(z)$ is the modified Bessel function of the second kind, satisfying $K_{i\nu}(z)=K_{-i\nu}(z)$.

When $k_\parallel<0$, the same procedure yields
\begin{eqnarray}\label{i3-2}
    I_{\pm}(\boldsymbol{k})=
    \frac{2}{a}\, e^{-i\frac{\omega_0\zeta}{a}} e^{\mp\frac{\pi\omega_0}{2a}} K_{{i\omega_0}/{a}}\left(\frac{k_\perp}{a}\right).
\end{eqnarray}
In both cases, the transition probabilities take the universal form
\begin{eqnarray}\label{p-3}
    |I_{\pm}(\boldsymbol{k})|^2=
    \frac{4}{a^2}\, e^{\mp\frac{\pi\omega_0}{a}} K^2_{{i\omega_0}/{a}}\left(\frac{k_\perp}{a}\right).
\end{eqnarray}
A striking consequence of Eq.~\eqref{p-3} is the emergence of acceleration-induced spectral blind spots. Both ordinary stimulated absorption and stimulated Unruh excitation are controlled by the same modified Bessel function $K_{{i\omega_0}/{a}}\left(\frac{k_\perp}{a}\right)$, which possesses real zeros. Therefore, whenever the ratio $k_\perp/a$ coincides with one of these zeros, both excitation channels vanish simultaneously. Consequently, the accelerated atom becomes effectively blind to incident photons at these discrete frequencies despite the presence of externally supplied radiation.

\emph{B. Collinear driving fields}---The spectral blind spots identified above arise only when the incident photon carries a transverse momentum component. To clarify the role of transverse momentum, we now consider the special case of collinear propagation. 
For photons propagating parallel to the acceleration, Eq.~\eqref{i1} simplifies to
\begin{eqnarray}\label{i1-1}
    I_{\pm}(\boldsymbol{k})=
    \int_{-\infty}^{\infty}d \tau\, e^{i\omega_0 \tau \mp i\frac{k_{0}}{a}e^{-a\tau}}.
\end{eqnarray}
Introducing the variable $\xi=(k_0/a)e^{-a\tau}$ and using standard integral identities gives~\cite{unruh1984happens,Takagi1986,Benjamin19}
\begin{eqnarray}\label{i1-2}
    I_{\pm}(\boldsymbol{k})=
    \frac{1}{a} e^{-i\frac{\omega_0}{a}\ln{\frac{a}{k_0}}} e^{\mp\frac{\pi\omega_0}{2a}} \Gamma(-i\omega_0/a).
\end{eqnarray}
Hence, the squared moduli $|I_{-}(\boldsymbol{k})|^2$ and $|I_{+}(\boldsymbol{k})|^2$, corresponding respectively to the probabilities of stimulated absorption and stimulated Unruh excitation for photons propagating along the acceleration are 
\begin{equation}\label{p-1}
    |I_{-}(\boldsymbol{k})|^2=\frac{2\pi}{a\omega_0}\left(1+\frac{1}{e^{2\pi\omega_0/a}-1}\right),
\end{equation}
and
\begin{equation}\label{p-2}
    |I_{+}(\boldsymbol{k})|^2=\frac{2\pi}{a\omega_0}\frac{1}{e^{2\pi\omega_0/a}-1}. 
\end{equation}
For antiparallel propagation, identical transition probabilities are obtained. 
Notably, these transition probabilities are independent of the photon momentum $\boldsymbol{k}$. Consequently, no blind spots occur in the collinear case. This result highlights the essential role of transverse momentum in generating the spectral blind spots.

\emph{Persistence under coherent-state driving}---So far, we have considered incident photons prepared in Fock states with sharply defined momentum. We now show that the spectral blind spots persist under coherent driving and can therefore be probed using realistic laser or microwave sources. To this end, we consider the transition amplitude from the initial state $|g,\alpha\rangle$ to the final state $|e,\beta\rangle$. Up to an irrelevant phase factor, it is given by~\cite{vsoda2022acceleration}
\begin{equation}\label{eq18}
    \mathcal{A}_{|g,\alpha\rangle \rightarrow |e,\beta\rangle}(\boldsymbol{k})
    =\lambda e^{-|\alpha-\beta|^2/2}(\alpha I_-(\boldsymbol{k})+\beta^* I_+(\boldsymbol{k})),
\end{equation}
where $|\alpha\rangle$ and $|\beta\rangle$ are coherent states satisfying $|\langle\alpha\beta\rangle|^2=e^{-|\alpha-\beta|^2}$, and $ I_{\pm}(\boldsymbol{k})$ are defined in Eq.~\eqref{I}. 
To obtain the excitation probability of the atom without conditioning on the final field state, we integrate over all possible final coherent states using the coherent-state resolution of identity. This gives
\begin{eqnarray}\label{eq19}
    \mathcal{P}_{\alpha}(\boldsymbol{k})
    &=&
    \int \frac{d^2\beta}{\pi}
    \left|
    \mathcal{A}_{|g,\alpha\rangle \rightarrow |e,\beta\rangle}(\boldsymbol{k})
    \right|^2
    \nonumber\\
    &\propto&
    \left|
    \alpha I_-(\boldsymbol{k})+\alpha^* I_+(\boldsymbol{k})
    \right|^2
    +
    \left|I_+(\boldsymbol{k})\right|^2 .
\end{eqnarray}
For a real coherent amplitude $\alpha$, this expression reduces to
\begin{equation}
    \mathcal{P}_{\alpha}(\boldsymbol{k})
    \propto
    |\alpha|^2
    \left|I_-(\boldsymbol{k})+I_+(\boldsymbol{k})\right|^2
    +
    \left|I_+(\boldsymbol{k})\right|^2 .
\end{equation}
The first term is the stimulated contribution, which scales with the photon number $|\alpha|^2$, while the second term represents spontaneous Unruh excitation. 
For noncollinear photons, both $I_{+}(\boldsymbol{k})$ and $I_{-}(\boldsymbol{k})$ are proportional to the same modified Bessel function,
$K_{i\omega_0/a}\left(\frac{k_\perp}{a}\right)$, 
and therefore share the same set of zeros. Consequently, the stimulated contribution to the excitation probability vanishes at these blind-spot frequencies, independently of the coherent-state amplitude. Thus, the predicted unresponsiveness is not restricted to Fock states, but persists for coherent incident fields. 
Importantly, Eq.~\eqref{eq19} is obtained after summing over all possible final field states. The blind spots thus appear directly as frequency-selective dips in the atomic excitation spectrum and can be observed without resolving or postselecting the outgoing radiation.

\emph{Discussion}---The central result of this work is the emergence of spectral blind spots in the stimulated response of uniformly accelerated atoms. At discrete frequencies, an accelerated atom becomes effectively insensitive to incident radiation even though photons capable of driving the transition are present. This behavior contrasts sharply with conventional resonance phenomena: whereas an inertial atom is characterized by an enhanced response at resonance, uniform acceleration generates well-defined frequencies at which stimulated transitions disappear. The resulting blind spots reveal an unexplored form of acceleration-modified light--matter interaction and provide a distinctive manifestation of the Unruh effect through the suppression, rather than the enhancement, of atomic transitions.

The origin of these blind spots is kinematic but genuinely non-inertial. Along a uniformly accelerated trajectory, the phase of an incident mode is no longer a simple sinusoidal function in the atom's proper time. Instead, the continuously varying Doppler factor reshapes the driving field and gives rise to the modified Bessel-function structure in Eq.~\eqref{p-3}. The same Bessel factor multiplies both $I_-(\boldsymbol{k})$ and $I_+(\boldsymbol{k})$, while their relative magnitude is fixed by the Unruh thermal factor, $|I_+(\boldsymbol{k})|^2=e^{-2\pi\omega_0/a}|I_-(\boldsymbol{k})|^2$. Consequently, both ordinary stimulated absorption and stimulated Unruh excitation vanish simultaneously at the blind-spot frequencies.

This perspective suggests a direct spectroscopic protocol. One irradiates the accelerated two-level atom with a tunable coherent field whose propagation direction has a transverse component relative to the acceleration, and monitors the atomic excitation probability while sweeping the incident frequency. Away from the blind spots, the stimulated signal scales with the photon number and can therefore be amplified by increasing the drive intensity. At the blind-spot frequencies, the stimulated contribution is strongly suppressed, leaving only weak spontaneous processes. The observable signature is therefore a frequency-selective dip in the atomic excitation spectrum rather than a small excess signal above background.

\begin{figure}
    \centering
    \includegraphics[width=0.9\linewidth]{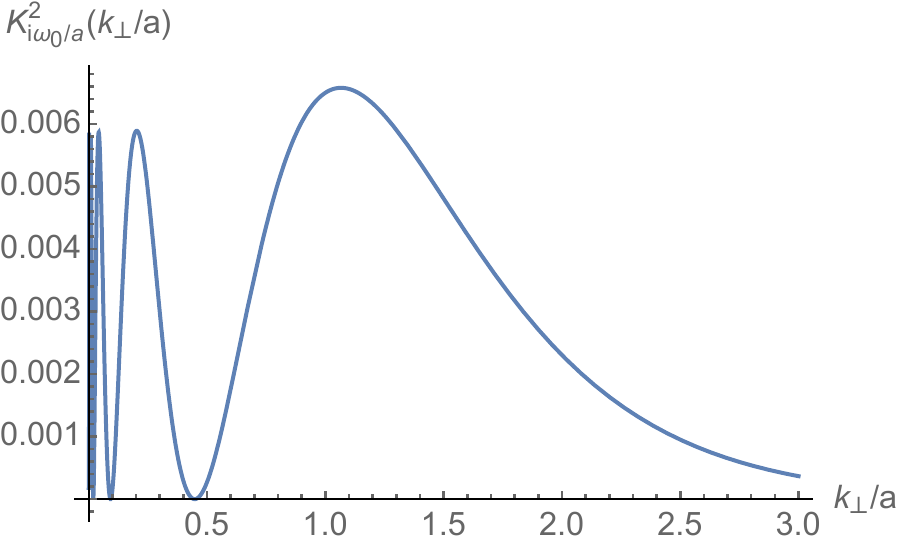}
    \caption{$K^2_{i\omega_0/a}(k_\perp/a)$ governing the transition probabilities for both stimulated absorption and stimulated Unruh excitation by transversely propagating photons. Here $\omega_0/a=2$.}
    \label{fig1}
\end{figure}

The spectral blind spots are robust against moderate parameter uncertainties and do not require unrealistically precise tuning. Figure~\ref{fig1} plots $K^2_{i\omega_0/a}(k_\perp/a)$, which governs the transition probabilities for both stimulated absorption and stimulated Unruh excitation for transversely propagating photons. The function varies smoothly near its zeros. For example, if $k_\perp/a$ differs by $10\%$ from the zero at $k_\perp/a\simeq0.448$, $K^2_{i\omega_0/a}(k_\perp/a)$ remains only $\sim4\%$ of its peak value. Small angular deviations primarily shift the effective value of $k_\perp/a$, and a finite spectral or angular width therefore broadens the dip rather than eliminating it.

The collinear limit further clarifies the role of acceleration-induced phase modulation. In the ideal infinite-time limit, photons propagating strictly parallel or antiparallel to the acceleration yield the frequency-independent probabilities in Eqs.~\eqref{p-1} and \eqref{p-2}. Physically, the Doppler-shifted driving frequency sweeps over an unbounded range during uniform acceleration, so every incident frequency becomes resonant at some proper time. The non-collinear result should therefore not be obtained by naively taking $k_\perp\to0$ in Eq.~\eqref{p-3}; the collinear limit is singular in the infinite-time idealization. For finite interaction times, however, numerical calculations show that the response approaches the collinear result continuously. Moreover, for noncollinear photons, the integrand becomes rapidly oscillatory at large $|\tau|$, so interaction times much longer than $1/a$ already provide an excellent approximation to the infinite-time blind spots.

It is useful to compare this mechanism with other acceleration-induced light-matter effects. For specially engineered nonuniform trajectories, stimulated absorption and stimulated Unruh excitation can have distinct zeros, allowing one process to be suppressed while the other remains finite \cite{vsoda2022acceleration}. In the uniformly accelerated case considered here, however, the two sets of zeros coincide. This simultaneous cancellation is tied to the canonical Unruh trajectory and yields a clean null signature in the total stimulated response, in contrast to engineered non-uniform trajectories that differ fundamentally from the uniform acceleration underlying the Unruh effect.

Finally, we comment on experimental feasibility. The scalar-field detector model used above is an idealized description that isolates the essential phase mechanism. A realistic electromagnetic implementation would require spin dynamics, polarization, finite beam geometry, field inhomogeneity, and radiation losses to be treated explicitly. As an order-of-magnitude feasibility assessment, one may use a free electron in combined electric and magnetic fields: the electric field supplies the acceleration, while a magnetic field aligned with the trajectory produces a tunable spin splitting. For a magnetic field of $1~\mathrm{T}$, the spin splitting corresponds to an angular-frequency gap $\omega_0\simeq1.8\times10^{11}~\mathrm{s}^{-1}$. Taking $\omega_0/a=2$, as in Fig.~\ref{fig1}, requires an acceleration of about $2.7\times10^{19}~\mathrm{m/s^2}$, comparable to accelerations of order $10^{20}~\mathrm{m/s^2}$ discussed for electrons in accelerator settings \cite{Bell83}. Radiation incident transverse to the trajectory with angular frequencies in the range $3.8\times10^9$--$9.6\times10^{10}~\mathrm{s}^{-1}$ would then include blind-spot frequencies near $8.3\times10^9$ and $4.0\times10^{10}~\mathrm{s}^{-1}$. Observing a driven-response dip at these frequencies would provide evidence for the predicted acceleration-induced spectral blind spots.

\emph{Summary}---We have shown that uniform acceleration can qualitatively reshape stimulated light--matter interaction by creating spectral blind spots in stimulated atomic transitions. For a uniformly accelerated two-level atom driven by obliquely incident photons, the transition amplitudes for ordinary stimulated absorption and stimulated Unruh excitation are controlled by the same modified Bessel function. At its zeros, both excitation processes vanish simultaneously, and the atom becomes effectively unresponsive to the incident field. We further showed that the blind spots persist under coherent-state driving and therefore remain observable with realistic laser or microwave sources. Rather than requiring postselection or measurement of the outgoing radiation, they appear directly as frequency-selective dips in the atomic excitation spectrum.

More broadly, our results reveal an unexplored regime of acceleration-modified light--matter interaction in which acceleration suppresses, rather than enhances, stimulated transitions. This mechanism provides a distinctive manifestation of the Unruh effect through the disappearance of driven atomic response and suggests a possible route toward experimental studies using high-acceleration electron platforms.

\begin{acknowledgments}
We would like to thank Yuebing Zhou for valuable discussions. 
This work was supported in part by the NSFC under Grants No. 12075084 and 12575051, and the innovative research group of Hunan Province under Grant No. 2024JJ1006.
\end{acknowledgments}

\end{document}